\documentclass[twocolumn,showpacs,amsfonts,aps,prc,nofootinbib,floatfix,%
superscriptaddress]{revtex4}

\usepackage{float}

\usepackage{amsmath}
\usepackage{bm}
\usepackage{graphicx}
\usepackage{xcolor}

\usepackage{epsfig}
\newcommand{\beq}{\begin{equation}}
\newcommand{\eeq}{\end{equation}}
\newcommand{\bea}{\vspace{0.25cm}\begin{eqnarray}}
\newcommand{\eea}{\end{eqnarray}}

\newcommand{\ro}{\mbox{{\boldmath
$\rho$}}}

\newcommand{\bb}{{{\bf b}}}

\def\lsim{\mathrel{\rlap{\lower4pt\hbox{\hskip1pt$\sim$}}
    \raise1pt\hbox{$<$}}}         
\def\gsim{\mathrel{\rlap{\lower4pt\hbox{\hskip1pt$\sim$}}
    \raise1pt\hbox{$>$}}}         

\begin{document}


\title{\Large\bf
  Nuclear modification factor  $I_{AA}$
  in $AA$ collisions at RHIC and LHC energies
  in scenarios with and without quark-gluon plasma
  formation in $pp$ collisions
}

\date{\today}

\author{B.G. Zakharov}

\address{
L.D.~Landau Institute for Theoretical Physics,
        GSP-1, 117940,\\ Kosygina Str. 2, 117334 Moscow, Russia
}

\begin{abstract}
We calculate the away-side hadron-triggered modification factor $I_{AA}$
in $AA$ collisions at RHIC and LHC energies for
scenarios with and without quark-gluon plasma formation in $pp$ collision.
We find that for both scenarios
theoretical results for $I_{AA}$ agree well with the available data
for 2.76 TeV Pb+Pb and 0.2 TeV Au+Au collisions.
We make predictions for $I_{AA}$ in 7 TeV O+O collisions
that are planned at the LHC. 
Our results show that measuring $I_{OO}$ in the whole
centrality interval and at small centrality ($\lsim 5$\%)
may give information on the presence of jet quenching
in $pp$ collisions.
\end{abstract}
%

\maketitle
\section{Introduction}
The experimental observation of the collective flow effects
and the strong suppression of high-$p_T$ hadrons 
in heavy ion collisions at RHIC and LHC provides a strong evidence of 
quark-gluon plasma (QGP) formation in $AA$ collisions
(for reviews see, e.g., \cite{hydro3,JQ_rev}).
The observation at LHC energies of the ridge effect
in $pp$ collisions
\cite{CMS_ridge,ATLAS_mbias} 
supports the idea \cite{Shuryak_QGP} that QGP formation is possible
in hadron collisions as well. 
The steep growth  of the strangeness  production in $pp$ collisions
at charge multiplicity density $dN_{ch}/d\eta\sim 5$ \cite{ALICE_strange}
also may be interpreted as strong evidence in favor of the QGP
formation in $pp$ events with sufficiently high multiplicity.
The scenario with the onset of the QGP formation regime in $pp$ collisions
at charge multiplicity density $dN_{ch}/d\eta\sim 5$ is also supported
by the analysis  of $\langle p_{T}\rangle$ as a function of
multiplicity \cite{Camp1}, employing Van Hove's arguments \cite{VH}.
If a liquid like droplet of the QGP is formed in $pp$ collisions, then,
similarly to $AA$ collisions,
there must be some jet modification (jet quenching) due to
radiative \cite{BDMPS1,LCPI1,W1,GLV1,AMY1} and collisional \cite{Bjorken1}
parton energy loss in the
QGP. 
In recent years the QGP formation and jet quenching in small
systems has received
considerable experimental and theoretical attention (see, e.g.,
a recent review \cite{UW_mQGP}).

The suppression of high-$p_T$ spectra in $AA$ collisions
is quantified by the nuclear modification factor $R_{AA}$.
It is defined through the particle yield in $AA$ collisions $N_{AA}$,
the number of events $N_{ev}$ and the $pp$ inclusive cross section
as
\beq
R_{AA}=\frac{d^2N_{AA}/dp_T^2dy}{N_{ev}
  \langle T_{AA}\rangle_{\Delta c} d^2\sigma_{pp}/dp_T^2dy},
\label{eq:10}
\eeq
where $\langle T_{AA}\rangle_{\Delta c}$ is the nuclear overlap function
for centrality bin $\Delta c$ (which is proportional to the number
of hard parton collisions $N_{col}$).
The factor $R_{AA}$ can be expressed through
the medium modification of the hard parton spectrum \cite{RAA_BDMS},
or alternatively via the medium modified
jet fragmentation functions (FFs) \cite{RAA08}. 
In the latter case, the $R_{AA}$ is dominated
by the behavior of the medium modified FFs
in the region of the fractional hadron transverse momentum $z_T$
close to unity.
The important feature of the $R_{AA}$ is that, for a given degree
of the jet modification in the QGP, its magnitude
decreases with increase of the slope of the $p_T$-dependence of hard
parton cross sections.

In the absence of the medium effects,
the difference between nuclear PDFs (nPDFs) and proton PDFs can lead to
a sizable deviation from unity of the theoretical nuclear
modification factor (we denote it by $R_{AA}^{pdf}$). The theoretical
uncertainties of $R_{AA}^{pdf}$, may make it difficult to observe
jet quenching for small systems, for which the effect of the jet modification
on $R_{AA}$ is weak, and may be of the same order as that from the nPDFs.
The theoretical predictions for $R_{AA}$ depend whether in $pp$ collisions the
QGP is formed or not, since in the scenario
with QGP formation in $pp$ collisions, the reference $pp$ hard inclusive
cross section in (\ref{eq:10}) differs from the pQCD one by the
medium modification
factor $R_{pp}$ \cite{Z_pp_PRL,Z_pp13} due to the medium jet modification
in the mini-QGP fireball formed in $pp$ collisions.
A detailed analysis of the RHIC and the LHC data on $R_{AA}$
in heavy ion collisions performed
in \cite{Z_hl}
 within the light-cone path integral (LCPI) approach to the induced
gluon emission \cite{LCPI1,LCPI2004} demonstrated that the data on
$R_{AA}$
in heavy ion
collisions can be describe equally well
in the scenarios with and without the QGP formation in $pp$ collisions.
The predictions for these two scenarios
begin to differ substantially for the light ion O+O collisions.
However, due to theoretical uncertainties of $R_{AA}^{pdf}$,
it may be difficult to discriminate
between the scenarios without and with QGP formation
from comparison with future LHC data
on the O+O collisions \cite{OO-CERN}.

An alternative method to probe the medium jet modification
is measuring the away-side factor $I_{AA}$ describing the
two-particle correlations
with the hadron/photon triggers \cite{Wang_di-h,PHENIX_di-h}.
Experimentally, the factor $I_{AA}$ is
defined as the ratio
\beq
I_{AA}=  \frac{Y_{AA}(\{p_T\},\{y\})}
{Y_{pp}(\{p_T\},\{y\})}\,,
\label{eq:20}
\eeq
where $Y_{AA}(Y_{pp})$ is 
the per-trigger particle ($h^t$) yield
  of the associated ($h^a$) hadron production
in $AA(pp)$ collisions, and 
$\{p_T\}=(p_T^{a},p_T^{t})$, $\{y\}=(y^{a},y^{t})$ are the sets of
the transverse momenta and rapidities
of the trigger particle and the associated hadron.
A unique advantage of the factor $I_{AA}$, as compared to the $R_{AA}$,
is that it is defined in terms of the self-normalized quantities
$Y_{AA,pp}$. For this reason,
for $I_{AA}$ there is no problem with the determination of the number
of hard process. 
Also, it is important that by measuring the away-side $I_{AA}$
at different $p^a$
one can probe the medium modification of the jet FFs in the broad range of
$z_T$. The factor $I_{AA}$ has a weaker sensitivity to the nPDFs than $R_{AA}$.
This makes it a good observable
to probe jet quenching in light ion collisions,
where the jet modification is weak.

The purpose of the present work is two-fold: (a) to address the question
whether the available data on $I_{AA}$ in heavy ion collisions
can be described within the jet quenching model of \cite{Z_hl}
with $\alpha_s$ obtained by
fitting the LHC heavy ion data on $R_{AA}$ for the scenarios
with and without QGP formation in $pp$ collisions, and (b) to obtain
theoretical predictions for $I_{AA}$ in O+O collisions
planned at the LHC in 2025 \cite{OO-CERN}.
The comparison with the heavy ion data on $I_{AA}$ is clearly necessary
step for understanding the robustness of the theoretical predictions
for O+O collisions. Of course, analysis of the heavy ion data
is interesting in itself, since comparison with data on $I_{AA}$
allows one to test the jet quenching scheme in a different region 
of the variable $z_T$ 
(for the available heavy ion data on $I_{AA}$
\cite{PHENIX24,ALICE11,ALICE16}
it is $z_T\sim 0.1-0.5$) as compared to the case of $R_{AA}$,
which is sensitive to the FFs at $z_T$ close to unity.

\section{Outline of the theoretical model}
\subsection{Per-trigger yields $Y_{AA,pp}$ in terms of
  di-hadron and one-hadron hard $NN$ cross sections}
For a given centrality, $c$,
the per-trigger yield $Y_{AA}$ for production of the trigger hadron
$h^t$ and the associated hadron $h^a$ in an $AA$ collision
can be written via the medium modified di-hadron (back-to-back for the
away-side $Y_{AA}$) and one-hadron
inclusive $NN$ cross sections as
\beq
Y_{AA}(\{p_T\},\{y\})=\left\langle
\frac{d^4\sigma_{NN}^m}{dp_T^adp_T^tdy^ady^t}\right\rangle_{AA}
\Bigg/\left\langle\frac{d^2\sigma_{NN}^m}{dp_T^tdy^t}\right\rangle_{AA}\,.
\label{eq:30}
\eeq
Here, $\langle \dots\rangle_{AA}$ refers to averaging over the geometry of 
the jet production in $AA$ collisions. For a given impact
parameter $\bb$, in terms of the jet
production transverse coordinate vector $\ro_j$, and the azimuthal
angle $\phi$ of the jet corresponding to the trigger particle,
$\langle \dots\rangle_{AA}$ for a function $F(\ro_j,\phi,\bb)$ is
defined as
\beq
\left\langle F \right\rangle_{AA}=
\frac{\int d\ro_j d\phi F(\ro_j,\phi,\bb) T_{AA}(\bb,\ro_j)}{2\pi\int d\ro_jT_{AA}(\bb,\ro_j)}\,,
\label{eq:40}
\eeq
where
$T_{AA}(\bb,\ro_j)=T_A(\ro_j)T(\ro_j-\bb)$
with $T_A(\ro)=\int dz n_A(\ro,z)$ the nuclear thickness profile
(here $n_A$ is nuclear density).

In the scenario without QGP formation in $pp$ collisions
one can use for the per-trigger yield $Y_{pp}$, that appears in the
denominator of (\ref{eq:20}), its value calculated in the pQCD, $Y_{pp}^{pQCD}$.
Then, the theoretical factor $I_{AA}$ can be written as 
\bea
I_{AA}^{th}(\{p_T\},\{y\})=\frac{
  \left\langle\frac{d^4\sigma_{NN}^m}{dp_T^adp_T^tdy^ady^t}\right\rangle_{AA}
  \Big/\left\langle\frac{d^2\sigma_{NN}^m}{dp_T^tdy^t}\right\rangle_{AA}}
{\frac{d^4\sigma_{pp}^{pQCD}}{dp_T^adp_T^tdy^ady^t}
  \Big/\frac{d^2\sigma_{pp}^{pQCD}}{dp_T^tdy^t}}
\,.
\label{eq:50}
\eea
In the scenario with QGP formation in $pp$ collisions,
$Y_{pp}$ includes the medium jet modification. Formally, it can be written
as $Y_{pp}=I_{pp}^{th}Y_{pp}^{pQCD}$, where $I_{pp}^{th}$ quantifies the
medium effect in $pp$ collisions. In this scenario, the right-hand side
of the formula (\ref{eq:50}) for $I_{AA}^{th}$ should be additionally
multiplied by the factor $1/I_{pp}$.
The formula for the $I_{pp}$ can be written as  (similarly
to (\ref{eq:50}) for $AA$ collisions)
\bea
I_{pp}^{th}(\{p_T\},\{y\})=\frac{
  \left\langle\frac{d^4\sigma_{pp}^m}{dp_T^adp_T^tdy^ady^t}\right\rangle_{pp}
  \Big/\left\langle\frac{d^2\sigma_{pp}^m}{dp_T^tdy^t}\right\rangle_{pp}}
{\frac{d^4\sigma_{pp}^{pQCD}}{dp_T^adp_T^tdy^ady^t}
  \Big/\frac{d^2\sigma_{pp}^{pQCD}}{dp_T^tdy^t}}
\,,
\label{eq:60}
\eea
where $\langle \dots\rangle_{pp}$ means averaging over the geometry of 
the jet production in $pp$ collisions.
The $\langle \dots\rangle_{pp}$ is defined similarly to
$\langle \dots\rangle_{AA}$
with the nuclear overlap
function $T_{AA}$ replaced by the overlap function for $pp$ collision.
In calculating $I_{pp}$ we perform averaging over the geometry of jet
production in $pp$ collisions for the central $pp$ collisions.
It is reasonable since we need the yield $Y_{pp}$ averaged over the
azimuthal angle. We calculate $T_{pp}$ for the gaussian parametrization
of the nucleon parton density in the transverse plane (we assume that
the quark and gluon distributions have similar form in the transverse
coordinates).   

Similarly to $R_{pp}$ \cite{Z_pp_PRL,Z_pp13}, the factor $I_{pp}$ for
the minimum bias $pp$ jet events is an unmeasurable quantity.
However, contrary to $R_{pp}$, due to the fact that $Y_{pp}$ is a
self-normalized quantity, it is possible to measure
the ratio between $I_{pp}$
for a given underlying event (UE) charge multiplicity and
the minimum bias $I_{pp}$
\cite{ALICE_Ipp,ALICE_Ipp_PLB}, which is equal to the ratio
$Y_{pp}/\langle Y_{pp}\rangle$. In the scenario with 
QGP formation this ratio should decrease with $dN_{ch}^{ue}/d\eta$
\cite{Z_pp_PRL,Z_Ipp}. 
In \cite{Z_Ipp} it was shown that the drop of the
ratio $I_{pp}/\langle I_{pp}\rangle$ with the UE charge multiplicity
observed by the ALICE collaboration in $5.02$ TeV $pp$ collisions
\cite{ALICE_Ipp,ALICE_Ipp_PLB}
is in reasonable agreement with calculations for the scenario with QGP
formation in $pp$ collisions.

\subsection{Jet quenching scheme for di-hadron and one-hadron cross sections}
The formulas for the medium modified di-hadron and one-hadron cross sections are
similar to that used in \cite{Z_Ipp,Z_IpA} for calculations of $I_{pp,pA}$
\bea
\frac{d^4\sigma_{NN}^m}{dp_T^adp_T^tdy^ady^t}
=\int \frac{dz^t}{z^t}D_{h^t/i}^m(z^t,p_{Ti})\nonumber\\
\times D_{h^a/j}^{m}(z^a,p_{Tj})
\frac{d^3\sigma_{ij}}{p_{Ti}dp_{Ti}dy_idy_j}\,,
\label{eq:70}
\eea
\beq
\frac{d^2\sigma_{NN}^m}{dp_T^tdy^t}
=\int \frac{dz^t}{z^t}D_{h^t/i}^{m}(z^t,p_{Ti})
\frac{d^2\sigma_{i}}{dp_{Ti}dy_i}\,.
\label{eq:80}
\eeq
Here
$\frac{d^3\sigma_{ij}}{dp_{Ti}dy_idy_j}$
is the cross section of $N+N\to i+j+X$ process for
$y_i=y^t$, $y_j=y^a$,
$p_{Ti}=p_{Tj}=p_{T}^t/z^t$ and $z^{a}=z^{t}p_T^a/p_T^t$, 
$\frac{d^2\sigma_{i}}{dp_{Ti}dy_i}$
is the cross section for $N+N\to i+X$ process,
$D_{h^t/i}^m$ and $D_{h^a/j}^m$
are the medium-modified FFs.
As in \cite{RAA08,RAA20,Z_hl}, $D_{h/i}^{m}$ is defined  
as a $z$-convolution 
\beq
D_{h/i}^{m}(Q)\approx D_{h/j}(Q_{0})
\otimes D_{j/k}^{in}\otimes D_{k/i}^{DGLAP}(Q)\,,
\label{eq:90}
\eeq
where $D_{k/i}^{DGLAP}$ is the DGLAP FF for $i\to k$ transition,
$D_{j/k}^{in}$ is the in-medium $j\to k$ FF,
and  $D_{h^{a,t}/j}$ are the FFs for hadronization transitions of the
parton $j$ to hadrons $h^{a,t}$.
The DGLAP FFs
$D_{k/i}^{DGLAP}$ are calculated  using the PYTHIA event generator \cite{PYTHIA}.
We use for FFs $D_{h/j}$ the KKP
\cite{KKP} parametrization  with $Q_0=2$ GeV.

The in-medium FFs $D_{j/k}^{in}$, which are a key ingredient
in calculating the medium-modified jet FFs, in our
jet quenching scheme are calculated 
through the induced gluon spectrum in the approximation
of the independent gluon emission \cite{RAA_BDMS}
supplemented by the momentum and the flavor sum rules
(we refer the interested reader to \cite{RAA20} for details).
The calculation of the induced
gluon spectrum is performed using the method of \cite{Z04_RAA}.
The collisional energy loss is calculated using the method of \cite{Z_coll}.
Its effect is treated as a perturbation to the radiative mechanism
(see \cite{Z_hl} for details).  
As in the analysis of \cite{Z_hl}, the induced gluon spectrum and the
collisional energy loss
are calculated with running $\alpha_s$ parametrized in the form
(supported by the lattice results for the in-medium $\alpha_s$
\cite{Bazavov_al1}) 
\beq
\alpha_s(Q,T)\, =\, \begin{cases}
\dfrac{4\pi}{9\log(\frac{Q^2}{\Lambda_{QCD}^2})}  & \mbox{if } Q > Q_{fr}(T)\;,\\
\alpha_{s}^{fr}(T) & \mbox{if }  Q_{fr}(T)\ge Q \ge cQ_{fr}(T)\;, \\
\frac{Q\alpha_{s}^{fr}(T)}{cQ_{fr}(T)} & \mbox{if }  Q < cQ_{fr}(T)\;, \\
\end{cases}
\label{eq:100}
\eeq
with $Q_{fr}=\kappa T$, $c=0.8$, $Q_{fr}(T)=\Lambda_{QCD}\exp\left\lbrace
{2\pi}/{9\alpha_{s}^{fr}(T)}\right\rbrace$ (we take $\Lambda_{QCD}=200$ MeV).
For the parameter $\kappa$ we take the values $2.5$ and $3.4$
for the scenarios with and without QGP formation in $pp$
collisions, respectively, fitted to 
the LHC data on the nuclear modification factor
$R_{AA}$ in heavy ion collisions.

\subsection{Model of QGP fireball}
We use the same model of the QGP fireball as used
in our previous global analysis of the data on $R_{AA}$ \cite{Z_hl}.
We describe the QGP evolution within Bjorken's 1+1D model
\cite{Bjorken}. It leads to the proper time
dependence of the entropy density
$s(\tau)/s(\tau_0)=\tau_0/\tau$ with $\tau_0$ the thermalization
time. At $\tau<\tau_{0}$ we take $s(\tau)=s(\tau_0)\tau/\tau_0$.
We set $\tau_{0}=0.5$ fm both for $AA$ and $pp$ collisions.
As in \cite{Z_hl}, we use a uniform fireball density distribution in
the transverse plane. The initial QGP entropy density for $AA$ collisions
is defined through the Bjorken relation \cite{Bjorken}
\beq
s_{0}=\frac{C}{\tau_{0} S_{f}}\frac{dN_{ch}(AA)}{d\eta}\,,
\label{eq:110}
\eeq
where $S_{f}$ is the overlap area of the colliding nuclei,
and $C=dS/dy{\Big/}dN_{ch}(AA)/d\eta\approx 7.67$ \cite{BM-entropy} 
is the entropy/multiplicity ratio. 
To calculate $dN_{ch}(AA)/d\eta$ we use the Glauber wounded  nucleon model
\cite{KN-Glauber} with parameters of the model as in our Monte-Carlo
Glauber analyses \cite{MCGL1,MCGL2,MCGL_Xe},
describing very well data on the midrapidity $dN_{ch}/d\eta$
for $0.2$ TeV Au+Au \cite{STAR02_Nch},
$2.76$ \cite{ALICE276_Nch} and $5.02$ TeV \cite{ALICE502_Nch} Pb+Pb, 
and $5.44$ TeV Xe+Xe \cite{ALICE544_Nch} collisions.
We use the Woods-Saxon nuclear density
$\rho_{A}(r)=\rho_{0}/[1+\exp((r-R_{A})/d)]$.
For Au(Pb) nucleus we take $R_A=6.37(6.62)$ and
$d=0.54(0.546)$ fm as in the GLISSANDO Glauber model \cite{GLISS2}
(in the PHOBOS Glauber model \cite{PHOBOS}).
For oxygen nuclear density we take $d=0.513$ fm \cite{ATDATA}
and $R_A=2.2$ fm (adjusted to have
$\langle r_{ch,O}^2\rangle=7.29$ fm$^2$ \cite{GLISS2}).
Our Glauber model calculations give for centralities
$\sim 5-10$\%
the ideal gas initial
temperatures $T_{0}\sim 400$ and $320$ MeV for 2.76 TeV Pb+Pb
and 0.2 TeV Au+Au collisions, respectively, and  for 7 TeV O+O collisions
we obtain $T_0\sim 280$ MeV.

To fix $T_{0}$ for $pp$ collisions,  we use the relation
(\ref{eq:110}) with  $dN_{ch}(AA)/d\eta$ replaced by the $pp$ UE
 charge multiplicity
density $dN_{ch}^{ue}(pp)/d\eta$, and
take $S_f=\pi R_f^2$, where $R_f$ is the effective radius of the mini-QGP
fireball in the $pp$ collision (corresponding
to an average radius for the whole range of the impact parameter).
We determine $R_f$
using the prediction for $R_{f}$ obtained 
in numerical simulations performed in \cite{glasma_pp} 
within the IP-Glasma model (see \cite{Z_hl} for details).
Using the experimental data on 
$dN_{ch}^{ue}(pp) /d\eta$ (see \cite{Z_hl} for details) we obtain
\beq
R_{f}[\sqrt{s}=0.2,2.76,7\,\, \mbox{TeV}]
\approx[1.26,1.44,1.51]\,\,\mbox{fm}\,.
\label{eq:120}
\eeq
Then,
from (\ref{eq:120}),
we obtain for the initial temperature of the QGP fireball
for the ideal gas entropy and for the lattice entropy \cite{t-lat}
(numbers in brackets)
\bea
T_{0}[\sqrt{s}=0.2,2.76,7\,\,\mbox{TeV}]
\approx\nonumber\\
\approx    [195(226),217(247),232(261)]\,\,\mbox{MeV}\,.
\label{eq:130}
\eea
Note that the possible theoretical uncertainties in the value of $R_f$ are not
important for are results, since the variation of $I_{pp}$ with $R_f$ is very
small. Similarly to the case of $R_{pp}$ \cite{Z_pp13,Z_hl},
this occurs due to a compensation 
between the enhancement of the energy loss caused by increase of the 
fireball size and its suppression caused by reduction of the fireball density.
\begin{figure} 
\begin{center}
\includegraphics[height=5.5cm]{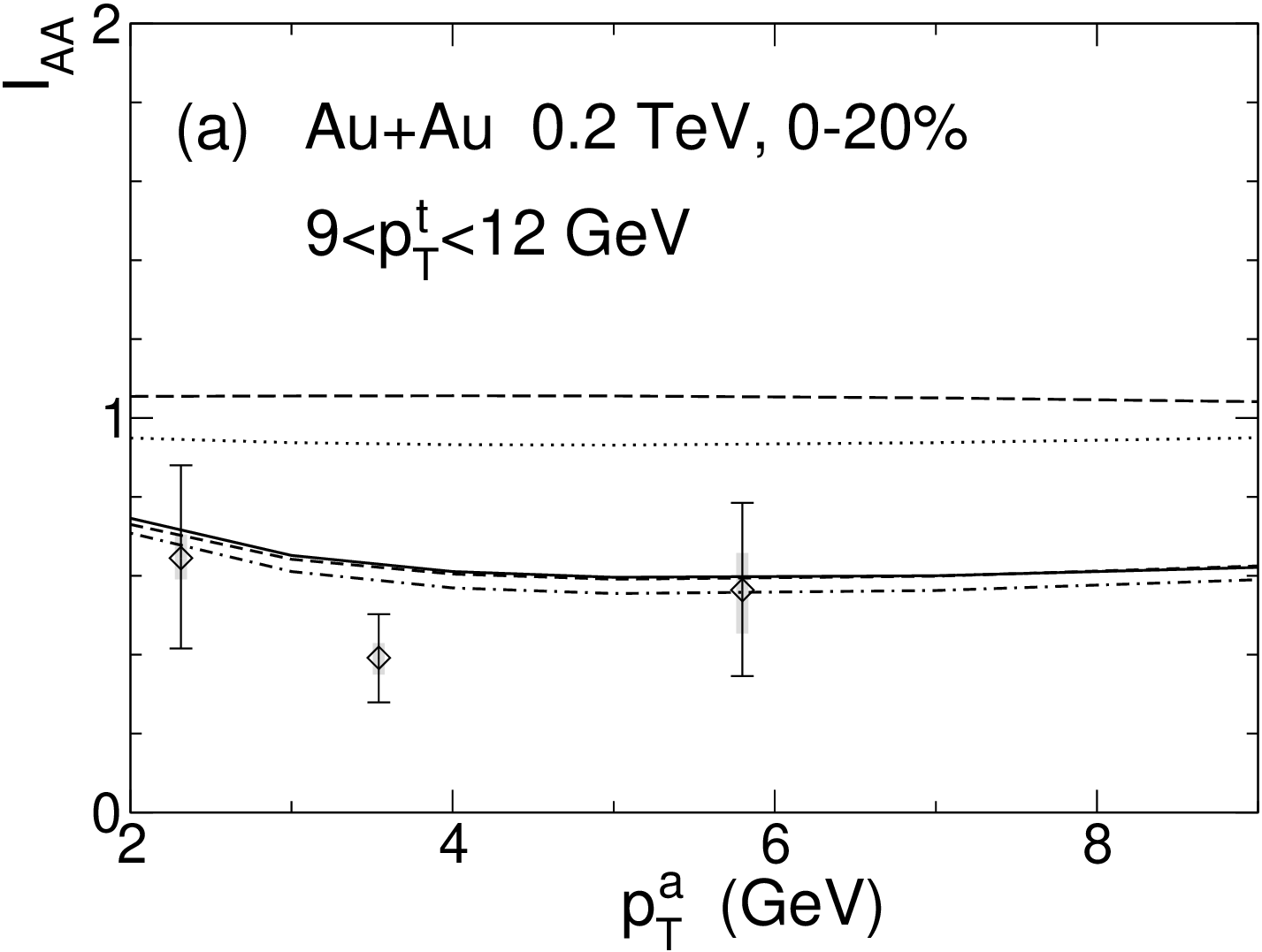}  
\includegraphics[height=5.5cm]{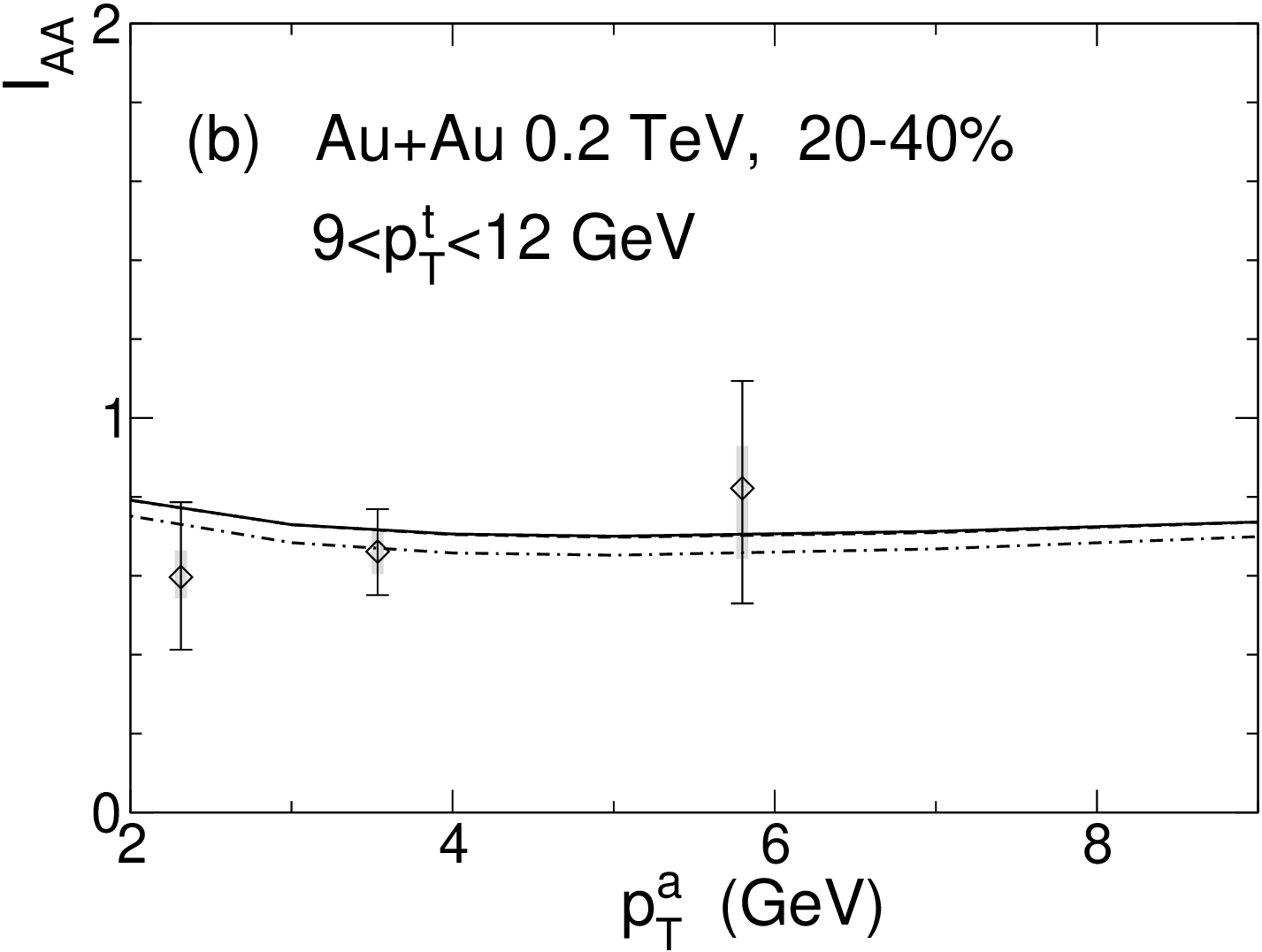}  
\end{center}
\caption[.]{The away-side $I_{AA}$ versus $p_{T}^{a}$ for 0.2 TeV Au+Au
collisions for (a) 0-20\% and
(b)  20-40\% centralities in the scenarios with (solid) and without
(dashed) QGP formation in $pp$ collisions.
Dot-dashed curves are obtained for
the intermediate scenario with QGP formation in $pp$ collisions only
at LHC energies.
Long-dashed curve shows $I_{AA}^{pdf}$, and dotted one shows $I_{pp}$.
Data points are from PHENIX \cite{PHENIX24}
for $\pi^0$ trigger.
 }
\end{figure}
\begin{figure} [!h] 
\begin{center}
\includegraphics[height=5.5cm]{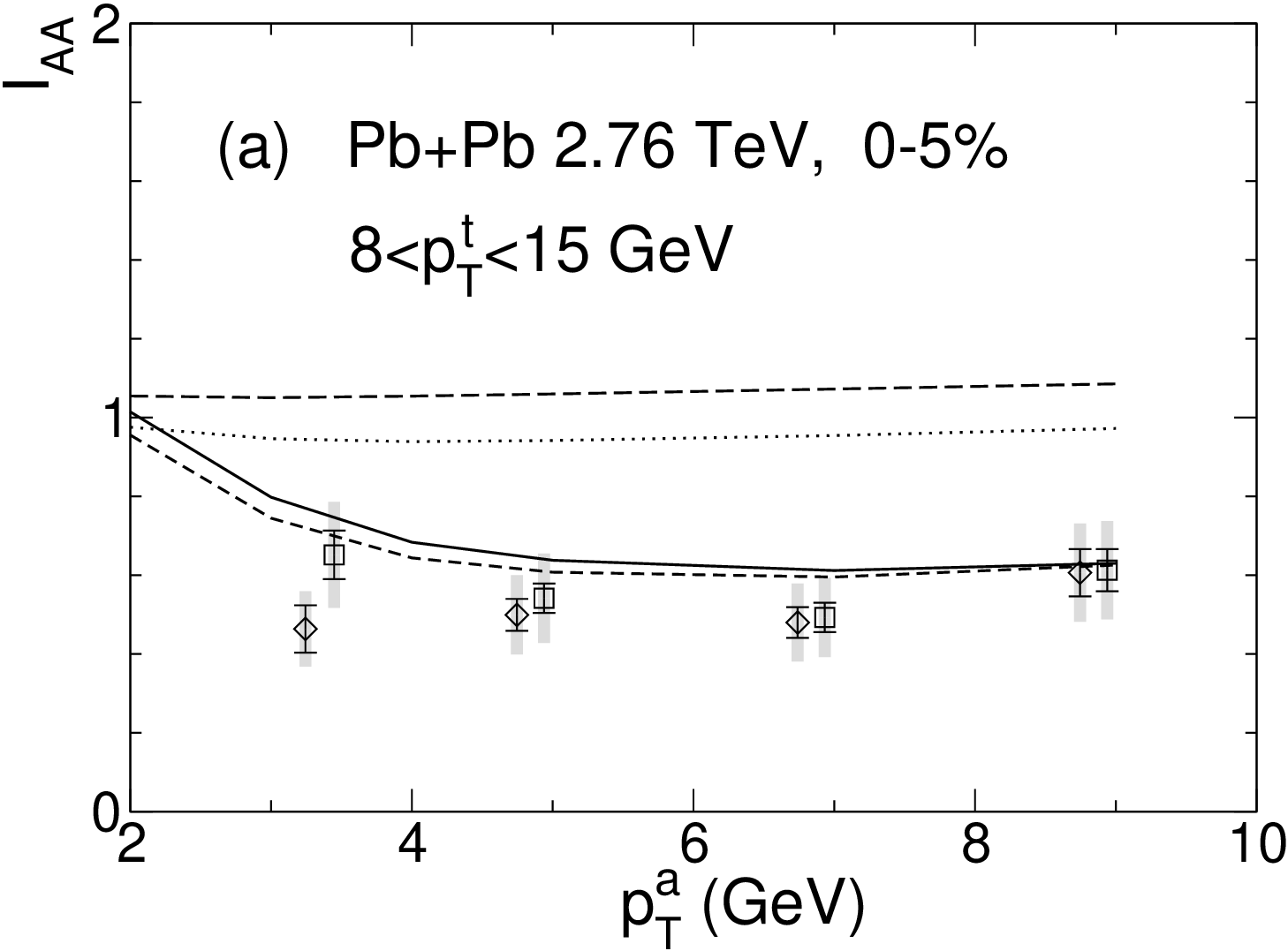}  
\includegraphics[height=5.5cm]{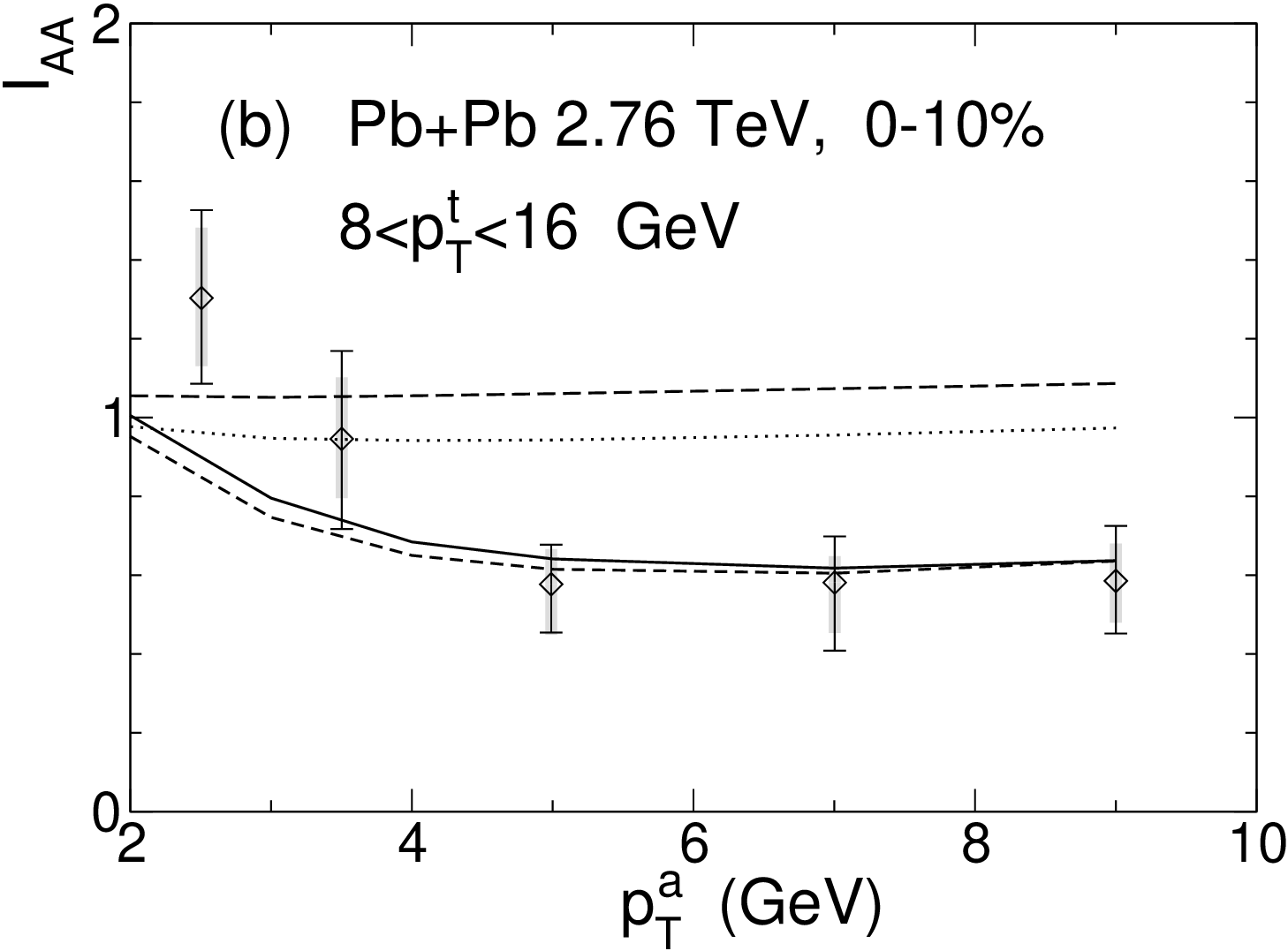}  
\includegraphics[height=5.5cm]{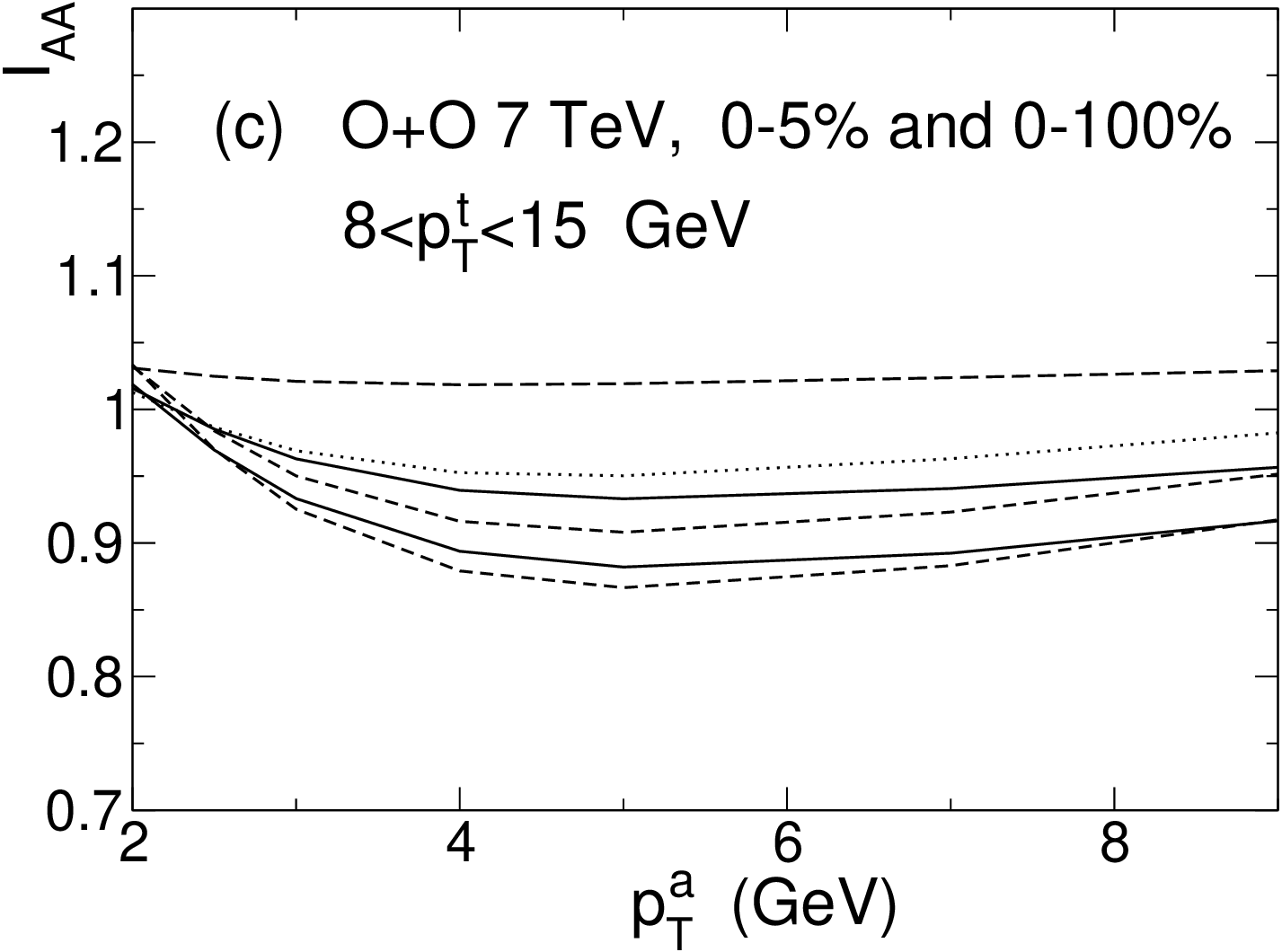}  
\end{center}
\caption[.]
{The away-side $I_{AA}$ versus $p_{T}^{a}$ in $AA$ collisions at LHC energies
in the scenarios with (solid) and without
(dashed) QGP formation in $pp$ collisions
in 2.76 TeV Pb+Pb collisions (a) for 0-5\% centrality and
          $8<p_T^{t}<15$ GeV, (b) for 0-10\% centrality and
$8<p_T^{t}<16$ GeV, (c) in 7 TeV O+O collisions
for (top to bottom) 0-100\% and 0-5\% centrality at $8<p_T^{tr}<15$ GeV.
Long-dashed curves show $I_{AA}^{pdf}$, and dotted curves show $I_{pp}$. 
Data points are (a) from ALICE  \cite{ALICE11}
(obtained using flat (squares) and $v_2$ (diamonds) backgrounds),
(b) from ALICE \cite{ALICE16}.
 }
\end{figure}
\section{Numerical results}
In this section we compare our calculations with the data on
$I_{AA}$
for 0.2 TeV Au+Au collisions
from PHENIX \cite{PHENIX24} and for 2.76 TeV Pb+Pb collisions from
the ALICE \cite{ALICE11,ALICE16}, and show
predictions for $I_{AA}$ in 7 TeV O+O collisions.
Besides the results for $I_{AA}$,  we also present results for unmeasurable
theoretical factors $I_{pp}$ and $I_{AA}^{pdf}$ (which
illustrates the effect of the difference between the nuclear and
the proton PDFs on $I_{AA}$).

In Fig. 1 we show our results for the $p_T^a$ dependence of
$I_{AA}$ in 0.2 TeV
Au+Au collisions for 0--20\% and 20--40\% centralities
for $9<p_T^t<12$ GeV and compare to recent data 
from PHENIX \cite{PHENIX24}.
We show $I_{AA}$ for the scenarios
with and without QGP formation in $pp$ collisions, and
for the intermediate scenario,
when QGP formation in $pp$ collisions occurs only at LHC energies.
The intermediate scenario is reasonable
since for $\sqrt{s}=0.2$ TeV the typical $pp$ UE charge midrapidity
multiplicity density is $\sim 5$ that is likely not to large
enough for a fully-fledged QGP formation regime (in the light of
the results of \cite{Camp1}).
From Fig. 1 one can see that for all
three scenarios theoretical results for $I_{AA}$ are close to each other,
and are in reasonable
agreement with the PHENIX data \cite{PHENIX24}.
This differs from the situation found in \cite{Z_hl} for the factor $R_{AA}$, 
for which the scenarios with and
without QGP formation in $pp$ collisions somewhat overshoot the PHENIX data \cite{PHENIX_r} at $p_T\sim 5-15$ GeV,
and the best agreement with the PHENIX data was found in the
intermediate scenario,
when QGP is formed only in $pp$ collisions at LHC energies.
From Fig. 1  we see that for $I_{AA}$, similarly to the case of
$R_{AA}$,
the difference between predictions 
for the scenarios with and without QGP formation in $pp$ collisions
is quite small. But, contrary to the case of $R_{AA}$,
the factor $I_{AA}$ for the intermediate scenario turns out to be close to
the predictions for scenarios with and without QGP formation in $pp$
collisions (both at RHIC and LHC energies). This occurs because
the deviation of $I_{pp}$ (dotted line in Fig. 1a) from unity is
considerably smaller than that for $R_{pp}$
($\sim 0.15-0.2$ at $p_T\sim 10$ GeV \cite{Z_hl}). 
As one can see from Fig. 1a for 0.2 TeV
Au+Au collisions deviation of $I_{AA}^{pdf}$ from unity
is rather small. This contrasts with $R_{AA}^{pdf}$,
for which the deviation from unity is as large as $\sim 15$\%
in the region $p_T\sim 5-20$ GeV \cite{Z_hl}.

In Fig. 2 we present results for $I_{AA}$ in 2.76 TeV Pb+Pb
and 7 TeV O+O collisions. 
In Figs. 2a and 2b we compare our results for $I_{AA}$
in 2.76 TeV Pb+Pb collisions to data from
ALICE \cite{ALICE11,ALICE16} 
for 0--5\% and 0--10\% centralities for the
trigger momentum windows $8<p_{T}^t<15$ and $8<p_{T}^t<16$ GeV.
As can be seen from Figs. 2a and 2b, our results for $I_{AA}$
in both the scenarios
are in reasonable agreement with the ALICE data. We see
that, similarly to 0.2 TeV Au+Au collisions, the difference between
predictions obtained in the scenarios
with and without QGP formation in $pp$ collisions is very small.
From Figs. 2a and 2b we observe that the deviation of $I_{AA}^{pdf}$ from unity
is quite small ($I_{AA}^{pdf}-1\sim 0.05-0.07$), and
$(1-I_{pp})\sim  0.03-0.06$.

In Fig. 2c we plot our results for $I_{AA}$ in 7 TeV O+O collisions
for  0-100\% (i.e., for the minimum bias O+O collisions)
and 0-5\% centrality classes for $8<p_{T}^t<15$ GeV.
From Fig. 2c we see that for the scenario with QGP formation
in $pp$ collisions the ratio
$|I_{AA}^{pdf}-1|/|I_{AA}^{pdf}-I_{AA}|$
is $\sim  0.25(0.15)$ for centrality class
0--100\% (0--5\%), and for the scenario without  QGP formation
in $pp$ collisions it is  $\sim  0.2(0.15)$
(in the region $p_T^a\gsim 4$ GeV). This says that it is
reasonable to expect that for 7 TeV O+O collisions the nPDFs
effects for $I_{OO}$ should be small compared to the effects of the
parton energy loss in the QGP.
From Fig. 2c one can see that for O+O collisions the ratio
${(1-I_{AA})_{0-5\%}}/{(1-I_{AA})_{0-100\%}}$
is noticeably different for the scenarios with and without QGP
formation in $pp$ collisions (at $p_T^a\sim 4-8$ it is about 1.9 and 1.4,
respectively).
For this reason, it may be used for experimental
discrimination between these two  scenarios.

Overall, our results agree quite well with the data on
$I_{AA}$ in heavy ion collisions \cite{PHENIX24,ALICE11,ALICE16}
for $p_T^a \sim 2-9$ GeV (that corresponds to $z_T\sim 0.1-0.5$).
Since in \cite{Z_hl}
we have obtained good description of the data on $R_{AA}$, which is sensitive
to the FFs at $z_T\gsim 0.5$, we can conclude that our jet quenching
scheme works quite well for $z_T\gsim 0.1$.
At first sight, it may seem somewhat strange for a model
which ignores the cascading induced gluon emission.
However, in fact, it is quite reasonable that the gluon cascading contribution
should not play a significant role
for hadrons with $p_T\gsim 2-3$ GeV.
Indeed, in heavy ion collisions
the dominant proper time region,
where the induced gluon emission can occur, is $\lsim 5$ fm.
For fast partons (say, with energy greater than a few tens of GeV)
the typical energy of the primary emitted soft gluons is
$\omega\sim 3-5$ GeV. The formation time/length for such gluons
is $\sim 4-6$ fm (see e.g. \cite{Z_pp13}). It means that, typically,
the secondary induced gluon emission should occur at the proper
time $\sim 5-10$ fm. But at such times the induced gluon emission rate
becomes small due to low density of the expanding QGP fireball.
For this reason the cascading processes should be of only marginal
significance in the jet modification for hadrons with
$p_T$ larger than a few GeV. 
For very soft hadrons with
$p_T^a \lsim 2$ GeV, the cascading induced gluon emission
with subsequent jet wake \cite{jet_wake1} hadronization may be important.

\section{Summary}
We have calculated the away-side hadron-triggered medium modification
factor $I_{AA}$ in $AA$ collisions at RHIC and LHC energies.
The medium modified FFs have been calculated within the LCPI approach to
induced gluon emission, treating the collisional energy loss as a perturbation.
We use a temperature dependent in-medium QCD running coupling $\alpha_s(Q,T)$
with a plateau around $Q\sim \kappa T$
(motivated by the lattice results \cite{Bazavov_al1}).
For scenarios with and without QGP formation in $pp$ collision, we
perform calculations of $I_{AA}$  without free parameters using
the values of $\kappa$ fitted to
the LHC data on the nuclear modification factor $R_{AA}$
in heavy ion collisions.
We found that, for both scenarios,
our theoretical results for $I_{AA}$ agree well with the data from
ALICE for 2.76 TeV Pb+Pb collisions \cite{ALICE11,ALICE16} and with
newly available data
from PHENIX for 0.2 TeV Au+Au collisions \cite{PHENIX24}.
Our results show that the difference in $I_{AA}$ between the scenarios
with and without QGP formation is very small for heavy ion collisions.

We make predictions for $I_{AA}$ in 7 TeV O+O collisions
that are planned at the LHC in 2025 \cite{OO-CERN}. 
Our calculations show that for O+O collisions the difference
in $I_{AA}$ between the scenarios with and without QGP formation is
sizeable. Our results show that measuring $I_{OO}$ in the whole
centrality interval and at small centrality ($\lsim 5$\%)
may give information on the presence of the medium jet modification
in $pp$ collisions.

\end{document}